\documentclass[aps,prl,showpacs,twocolumn,floats,superscriptaddress]
{revtex4}
\usepackage{amssymb,amsbsy,amsmath}
\usepackage[dvipdfmx]{graphicx}
\usepackage{bm}
\usepackage{color}
\makeatletter
\def\lsim{\mathrel{\mathpalette\gl@align<}}
\def\gsim{\mathrel{\mathpalette\gl@align>}}
\def\gl@align#1#2{\lower.6ex\vbox
{\baselineskip\z@skip\lineskip\z@
\ialign{$\m@th#1\hfil##\hfil$\crcr#2\crcr\sim\crcr}}}

\makeatother

\begin{document}

\title{Universal scaling in quenches across a discontinuity critical point}

\author{Sei Suzuki}
\affiliation{Department of Liberal Arts, Saitama Medical
University, Moroyama, Saitama 350-0495, Japan}

\author{Amit Dutta}
\affiliation{Department of Physics, Indian Institute of
Technology, Kanpur 208 016, India}

\date{\today}

\begin{abstract}
We study  slow variation (both spatial as well as temporal)  of a parameter of a system in the vicinity of  discontinuous quantum  phase transitions, in particular,
a discontinuity critical point (DCP) (or a first-order critical point).
We obtain the universal scaling relations of the density of defects
and the residual energy after a temporal quench, while we also unravel the scaling of the characteristic length scale
associated with  a spatial quench of a symmetry breaking
field. Considering a spin-1/2 XXZ  chain we establish  how
 these scaling relations get modified when the DCP is
located at the boundary of
a gapless critical phase; these predictions are also confirmed numerically.
\end{abstract}


\maketitle

When a macroscopic system is driven from a disordered phase to an
ordered phase by lowering the temperature across a phase transition, 
there is  spontaneous generation of topological defects, namely,  vortices, kinks and domain walls; this  has been explained invoking upon  the so called Kibble-Zurek (KZ) mechanism.
Originally  proposed in the context of
the cosmic evolution \cite{Kibble1976,Kibble1980} and following the subsequent generalization to the condensed matter systems
 \cite{Zurek1985}, in recent years the KZ mechanism has been extended to
the quenches in the vicinity of a quantum phase transition \cite{Zurek2005,Polkovnikov2005,Dziarmaga2005,damski05}.
In fact, the remarkable progress in  experiments
with cold atoms in optical lattices, which can be treated as nearly  isolated systems, 
has paved the way for this recent upsurge in theoretical studies of non-equilibrium dynamics of closed quantum systems.

A number of studies 
have established  that, when a $d$-dimensional quantum system is driven across an isolated quantum critical point (QCP), by changing a parameter of the Hamiltonian
in a linear fashion as $t/t_Q$, the density of defects ($\hat \rho$) satisfies the KZ scaling
 $t_Q^{-d\nu/(z\nu + 1)}$. Similarly when quenched to the QCP,  the residual energy density, i. e., the excess energy density over the ground state of the final Hamiltonian ($\varepsilon_{\rm res}$)
scales as $t_Q^{-(d + z)\nu/(z\nu + 1)}$; here,
$\nu$ is the critical exponent
of the correlation length, and $z$ is the dynamical critical
exponent associated with the corresponding QCP. On the contrary, when quenched to the gapped phase, the residual
energy follows a scaling relation identical to that of the defect.
These scaling formulae
 however get modified, sometimes drastically,  in several interesting situations 
\cite{Mukherjee2007,Divakaran2009,Divakaran2008,Sengupta2008,Dutta2010,Hikichi2010}. (see for review \cite{PolkovnikovRev,dutta10,dziarmaga10}).


The concept of the KZ scaling has also been extended to 
spatial quenches \cite{Platini2007,Zurek2008},
where a parameter is inhomogeneous
along a spatial coordinate $x$ characterised by  a slope $x_Q^{-1}$ with 
the quantum (or classical) phase transition 
occurring at $x=0$. The order parameter is found to  bend
smoothly over the region $|x|\lsim \hat{\xi_s}\sim x_Q^{\nu/(1 + \nu)}$; obviously, $\hat \xi_s$ sets a length scale characterized
by quantum critical exponents.
These predictions has been confirmed analytically 
 as well as numerically in 
 several models  \cite{Platini2007,Zurek2008,Damski2009}.

The present Letter studies temporal as well as spatial
quenches near a quantum \textit{discontinuous (first-order) phase transition}, which,
to the best
of our knowledge, has not received much attention in the context of quenching and the consequent KZ mechanism
scenario; however,
the time evolution across a discontinuous phase transition 
has been widely studied from the viewpoint of quantum annealing (or adiabatic quantum computation) mainly concentrating
on the scaling of the energy gap \cite{young10,jorg10}.
Although quite recently, Bonati \textit{et al.} \cite{Bonati2014}
proposed a scaling of the length scale $\hat{\xi}$
near a classical discontinuous transition in the presence of
a temperature gradient,   the same 
for  a  \textit{sloped}  symmetry breaking field is not known.

The goal of the present Letter is to show that
$\hat{\rho}$ and $\varepsilon_{\rm res}$
after a temporal quench across (or at) \textit{a discontinuity critical point} (DCP) (or \textit{a first-order critical point} \cite{fisher82, nishimori11}) do indeed follow universal scaling laws, given in terms of the associated critical exponents derived here for a generic DCP.
On the contrary, when a symmetry breaking field is spatially quenched
near the DCP, a characteristic length scale is also shown to emerge.
Even though similar scaling relations can be derived for classical quenches,
 our focus will be restricted to quantum quenches only.

\begin{figure}[t]
\begin{center}
\includegraphics[width=6cm]{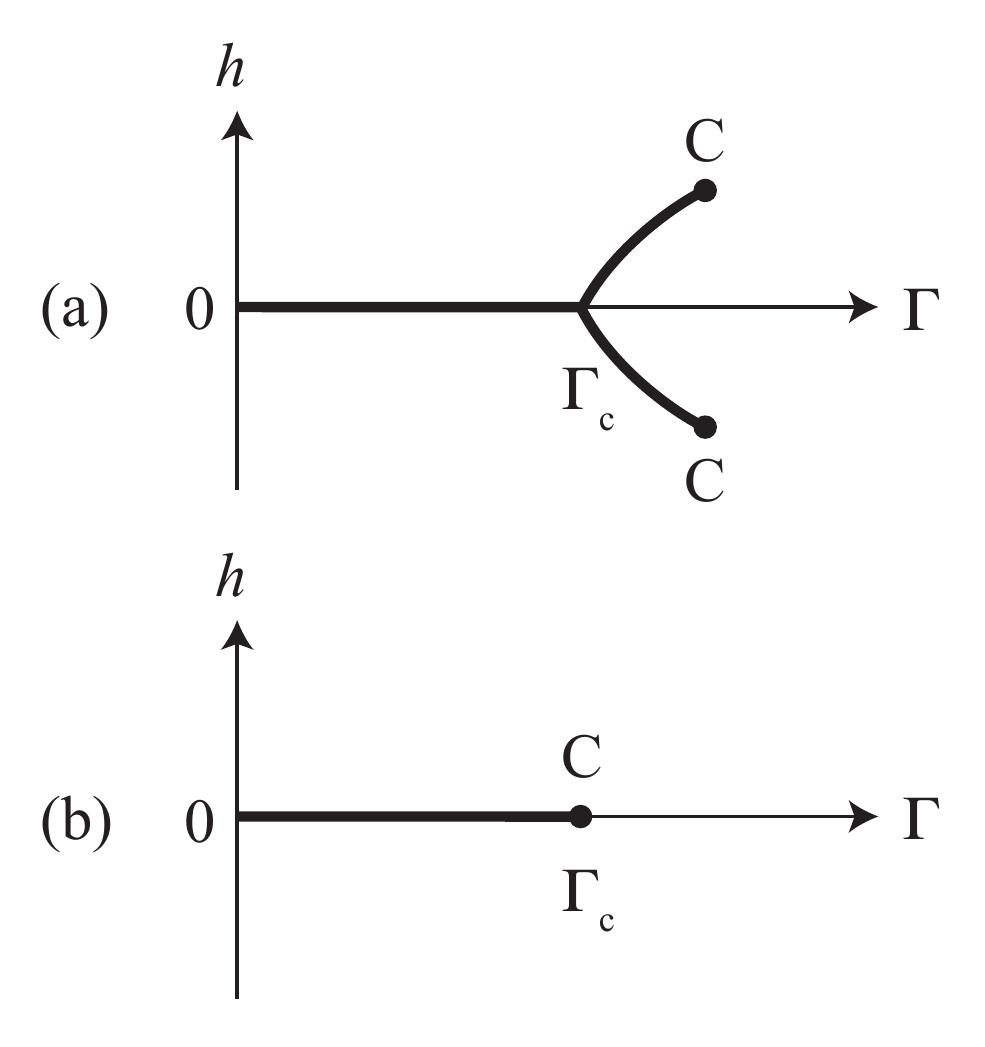}
\end{center}
\caption{Schematics of two phase diagrams involving a discontinuous
transition.}
\label{SDFig1}
\end{figure}

We initiate our discussion with  a phenomenological description
of discontinuous transitions, illustrating with the example of simple
Ising-like systems with $Z_2$ symmetry exposed to a tunable quantum fluctuation with its strength denoted by $\Gamma$.
Due to Ising symmetry,
doubly degenerate ordered ground states coexist at a low $\Gamma$
in the absence of a symmetry breaking field.
With increasing $\Gamma$, the order parameter characterizing the ordered states
vanishes discontinuously at a discontinuous transition 
$\Gamma_{\rm c}$. 
Application of a symmetry breaking field $h$, on the other hand, lifts  the degeneracy of two states with nonzero order parameter; as
$h$   is tuned across the zero-field coexisting line (see Fig.~1),
the order parameter shows a jump from one to the other. One may now wonder  what is the fate of 
this coexisting line beyond $\Gamma_{\rm c}$?
There are indeed two possibilities: in the former, the coexisting
line splits into two lines (Fig.~\ref{SDFig1}(a)) when the  order parameter shows 
a discontinuous jump to a finite value even at $\Gamma > \Gamma_{\rm c}$
with increasing the magnitude of $h$ due
to an interchange of a stable and a metastable states.
The coexisting line ends at a critical point denoted by C in 
Fig.~\ref{SDFig1}(a), beyond which metastable
states disappear.
This scenario can be explained within  the framework of the phenomenological Ginzburg-Landau-Wilson theory.
In the latter case, on the contrary, the coexisting line terminates at $\Gamma_{\rm c}$
(Fig.~\ref{SDFig1}(b)) with no metastable state appearing  for $\Gamma > \Gamma_c$.
 The transition point $\Gamma_{\rm c}$ is referred to as a DCP
 and is the focus point
of the present Letter.

To enunciate the  critical properties associated  with a DCP, one proceeds with the  following 
assumptions: at the DCP,
the ground-state energy density $\varepsilon(\Gamma, h)$
is a continuous function of $\Gamma$ and $h$, but its
derivatives $\partial \varepsilon(\Gamma, 0)/\partial \Gamma$ and
the order parameter $m(\Gamma, 0)$
 (at $h=0$) exhibit jumps  as a
function of $\Gamma$. We put them in 
a precise mathematical form:
\begin{eqnarray}
&& \varepsilon(\Gamma_{\rm c}+0, 0) = \varepsilon(\Gamma_{\rm c}-0, 0) 
\label{eq:Assump-e1}\\
&& \frac{\partial \varepsilon(\Gamma_{\rm c}+0, 0)}{\partial \Gamma} \neq 
 \frac{\partial \varepsilon(\Gamma_{\rm c}-0, 0)}{\partial \Gamma}, 
\label{eq:Assump-e2}\\
&& |m(\Gamma_{\rm c}-0, 0)| > m(\Gamma_{\rm c}+0, 0) = 0.
\label{eq:Assump-m1}
\end{eqnarray}
On the other hand, right at  $\Gamma = \Gamma_{\rm c}$, $m$
shows a discontinuous change with $h$ at $h = 0$:
\begin{equation}
 m(\Gamma_{\rm c}, + 0) \neq m(\Gamma_{\rm c}, - 0) . 
\label{eq:Assump-m2}
\end{equation}
Equipped with the above assumptions, let us now propose the  scaling
ansatz, first fixing $h=0$:
As usual,  the correlation length ($\xi$) and 
the ground-state energy density near a quantum critical point
follow scaling forms,
$\xi\approx A^{(\pm)}|\Gamma-\Gamma_{\rm c}|^{-\nu}$
and 
$\varepsilon(\Gamma, 0) - \varepsilon(\Gamma_{\rm c}, 0)
\approx B^{(\pm)}\xi^{-(d + z)}$ \cite{Sachdev, suzuki13}, 
where $\nu$ and $z$ are the correlation length and the dynamical
exponents respectively and $\pm$ stands for the sign of $\Gamma -
\Gamma_{\rm c}$. The first order derivative of $\varepsilon(\Gamma, 0)$
is written as $\partial \varepsilon(\Gamma,0)/\partial\Gamma\approx
B^{(\pm)}\xi^{-(d+z)+1/\nu}$.
This along with Eqs.~(\ref{eq:Assump-e1}) and (\ref{eq:Assump-e2}) leads to
$\nu = 1/(d + z)$.
In  the other situation, when 
$\Gamma = \Gamma_{\rm c}$ and $h \neq 0$,
the corresponding correlation length ($\xi_{\rm H}$) and the ground-state energy density
satisfy $\xi_{\rm H}\approx A^{(\pm)}_{\rm H}|h|^{-\nu_{\rm H}}$
and $\varepsilon(\Gamma_{\rm c}, h) - \varepsilon(\Gamma_{\rm c},
0)\approx B^{(\pm)}_{\rm H}\xi_{\rm H}^{-(d+z)}$.
Recalling that $m$ is determined by
$m(\Gamma, h) = -\partial \varepsilon(\Gamma, h)/\partial h$,
these relations together with Eq.~(\ref{eq:Assump-m2}) yields
$\nu_{\rm H} = 1/(d + z)$.

Henceforth, we move on to the non-equilibrium situation
arising due to quantum
quenches. 
We consider a temporal quench of the parameter field (with $h=0$) of the form
$\Gamma(t) = \Gamma_{\rm c} - \Gamma_0 \frac{t}{t_Q}$;
here $t$ is the time, $t_Q^{-1}$ is a quench rate,
and $\Gamma_0$ is a positive constant. Consequently, the system is driven from the 
the disordered phase (for $t < 0$)
to the ordered phase (for $t > 0$).
Assuming a  power-law scaling of 
the healing time
$\tau(T, 0)\sim \xi(T, 0)^z$,  
for a time-varying parameter, we introduce the notion of 
\textit{the instantaneous healing time}
$\tau(t) = \xi(\Gamma (t), 0)^z$; using 
$\xi(\Gamma, 0)\sim |\Gamma-\Gamma_{\rm c}|^{-\nu}$ and the quenching form
of $\Gamma$, this
can be written in the form
$\tau(t)\sim |t/t_Q|^{-z\nu}$.
If the system is initially prepared in the ground state, 
it remains in
the instantaneous ground state as long as the healing time is
short i.e.,  the system is far away from the QCP at $t=0$. This adiabatic time evolution breaks down,
when $t \to 0$  as a consequence of the diverging
healing time. 
In order to derive the scaling of the healing length, we invoke upon the
adiabatic-impulse approximation \cite{damski05} which assumes that
the the system is in the instantaneous ground state when 
$t < -\tau(t)$, while it freezes when $t > -\tau(t)$.  The  
 freezing time $\hat{t}$ is then determined by the condition
$|\hat{t}| = \tau(\hat{t})$.
One then immediately finds
$\hat{t}\sim t_Q^{z\nu/(z\nu + 1)} = t_Q^{z/(d + 2 z)}$.
For $t > \hat t$, the system  has an inhomogeneity
which is characterized by a length scale $\hat{\xi}$, which can be
derived form the scaling relation
\begin{equation}
 \hat{\xi}_{\rm t} \approx \xi(\Gamma (\hat{t}), 0) \sim t_Q^{1/(d + 2z)} .
\label{eq:xi_t}
\end{equation}
Noting that the defects are distributed in a $d$ dimensional phase space, the scaling of the density of defects 
is then given by $\hat{\rho}\sim \hat{\xi}_t^{-d}$, which yields
\begin{equation}
 \hat{\rho} \sim t_Q^{-d/(d + 2z)} .
\label{eq:rho}
\end{equation}
If the elementary excitation in the system after a temporal quench
has a gapless dispersion as $|\textbf{k}|^{z}$ with the wave vector
$\textbf{k}$, the residual-energy density  follows the scaling relation
\begin{equation}
 \varepsilon_{\rm res} \sim t_Q^{-(d + z)/(d + 2z)};
\label{eq:epsilon1}
\end{equation}
otherwise, it satisfies the scaling  given in Eq.~(\ref{eq:rho}).

In a similar spirit, let us now consider a spatial  quench of a parameter of the Hamiltonian, $ h(x) = h_0 (x/x_Q)$, keeping
$\Gamma$ fixed at $\Gamma_{\rm c}$,
where $x$ refers to one of the $d$ spatial coordinates, 
$x_Q^{-1}$ is the quench rate or the scale of inhomogeneity, and $h_0$ is a positive constant.
Note that the system is critical when the field vanishes, \textit{i.e.},  at around $x=0$.
Defining   \textit{the local correlation length}\cite{Zurek2008}
$\xi(x) = \xi(\Gamma_{\rm c}, h(x))$ and using the relation
$\xi(\Gamma_{\rm c}, h)\sim |h|^{-\nu_H}$ ($\nu_{\rm
H}=1/(d+z)$), one readily finds 
$\xi(x)\sim (|x|/x_Q)^{-\nu_{\rm H}}$.
In spite of the inhomogeneity, 
the system should maintain the local equilibrium 
for $x$ sufficiently far from $x=0$ since
the local correlation length becomes is much smaller than $x_Q$.
On approaching the critical situation, however, the local correlation
grows and hence the approximation of the  local equilibrium
breaks down  leading to a finite  correlation length
near $x=0$. To make an estimate of this finite
correlation length, one again invokes upon  the pulse-impulse
approximation: the system maintains its local equilibrium for
$x$ satisfying $|x| \gg \xi(x)$, while for $|x| \ll \xi(x)$ 
the system \textit{freezes} with the correlation length
$\hat{\xi}$. The boundary $\hat{x}$ of these two regimes
is approximately determined by the condition $\xi(\hat{x}) = |\hat{x}|$ and $\hat{\xi}$ is
given by $\hat{\xi_{\rm s}} = \xi(\hat{x})$ which results in the scaling relation
 $|\hat{x}| \sim x_Q^{\nu_{\rm H}/(1+\nu_{\rm H})} = x_Q^{1/(d+z+1)}$.
One therefore arrives at  the scaling relation of the length scale
\begin{equation}
 \hat{\xi}_{\rm s} \sim x_Q^{1/(d+z+1)} .
\label{eq:xi_s}
\end{equation}


%
%
Equations (\ref {eq:xi_t}),
(\ref{eq:epsilon1}) and (\ref{eq:xi_s}) constitute the main result of our paper regarding
the quenching in the vicinity of a generic DCP as depicted in Fig 1(b).
We
generalize these relations below  by exploring the dynamics of a  XXZ chain
in a magnetic field described by the Hamiltonian
\begin{equation}
 H = \sum_{i=1}^{L-1}
  \left(\sigma_i^x \sigma_{i+1}^x + \sigma_i^y \sigma_{i+1}^y
   + \Delta\sigma_i^x\sigma_{i+1}^z \right) - \sum_{i=1}^L h_i\sigma_i^z,
\label{eq:H_XXZ}
\end{equation}
where $\sigma_i^{\alpha}$'s ($\alpha = x, y, z$) are the Pauli operators
and $L$ is the system size. We first set
$h_i = 0, \forall i$,
 when the model exhibits a discontinuous quantum phase
transition at $\Delta = -1$. The ground state for $\Delta < -1$
possesses the complete ferromagnetic order, while it is a disordered
state for $-1 < \Delta < 1$.
Starting from the the ground state of the initial Hamiltonian with $\Delta=0$, we consider  a slow ramp to $\Delta=-1$ following a
linear protocol $\Delta (t) = -t/t_Q$.
Since the Hamiltonian commutes with 
$\sigma_{\rm tot}^z = \sum_{i=1}^L\sigma_i^z$ and
the initial state lies in the
$\sigma_{\rm tot}^z = 0$ sector,
the dynamics is restricted within this manifold  only.
Even then the transition occurring at $\Delta = \Delta_{\rm c}= -1$  \textit{indeed} can be viewed
as a DCP as we have established rigorously in the supplemental section. However, this DCP occurs at the end of a gapless critical line, which
results in modification of associated critical exponents and consequently, of the scaling relations of $\hat{\rho}$ and $\varepsilon_{\rm res}$.

The low-energy states of the XXZ chain 
with $-1 < \Delta \leq 1$ are described by the Tomonaga-Luttinger
liquid with gapless excitations \cite{Giamarchi,bib:Cabra}; this gaplessness, 
 rules out the possibility of defining the exponent $\nu$ 
using the conventional scaling 
of the energy gap 
$\mathit{\Delta} E\sim (\Delta - \Delta_{\rm c})^{z\nu}$
\cite{Sachdev}.
However, the dispersion relation at $\Delta_c$
is quadratic, i.e., energy gap $\mathit{\Delta} E \sim L^{-2}$  implying
 $z= 2$.

To estimate the exponent $\nu$, let us recall that the ground-state energy density in the
 $\sigma_{\rm tot}^z = 0$ manifold behaves as
 $\varepsilon(\Delta) - \varepsilon(\Delta_{\rm c}) = -
 (\Delta_{\rm c} - \Delta)$
 for $\Delta < -1$
 and $ \approx (\Delta - \Delta_{\rm c})^{3/2}$
 for $\Delta > -1$ \cite{yang66}. Comparing with the conventional scaling of the same, namely,
$(\Delta -\Delta_{\rm c})^{\nu(d+z)}$ along with $d=1, z=2$, one finds that $\nu=1/2$ which  clearly differs from
the predicted value $\nu=1/(d+z)$. This mismatch stems from the
 existence of the gapless line for $\Delta >\Delta_{\rm c}$
characterized by gapless excitations with dispersion $|k|^{z_+}$, with
 $z_+ = 1$ in the present case.

To settle this anomaly, we propose the following modified scaling forms
of the spectrum $\mathit{\Delta}E\sim |\Delta -
\Delta_{\rm c}|^{\theta_{\pm}}k^{z_{\pm}} + k^z$
and the ground-state energy density 
$\varepsilon(\Delta) -
\varepsilon(\Delta_{{\rm c}}) \sim
| \Delta - \Delta_{{\rm c}} |^{ \nu z }(1 +
 A_{\pm} |\Delta - \Delta_{\rm c}|^{-\nu})^{-d}$, where $z_{\pm}$ characterize the excitations on either sides of the DCP
 respectively; $z_{+}=1$ and $z_{-}=0$ (on the gapped side), in the present
 case. Similarly, $A_{\pm}$ are the critical amplitudes
(of $\xi$) 
on the gapless and the gapped sides. The case 
 $A_- = 0$ implies a perfectly ordered  ground state 
without any fluctuation; this is indeed  true for
 the XXZ chain which can be established using the same scaling of the ground-state energy density and the gap.

In order to derive the critical exponents of the DCP  associated with a gapless phase, it is convenient to introduce a quantity
$\phi(\Delta) = (\varepsilon(\Delta) - \varepsilon(\Delta_{\rm
c}))|\Delta - \Delta_{\rm c}|^{-\nu z_{\pm}}$ with $z_-$ for $\Delta <
\Delta_{\rm c}$ and $z_+$ for $\Delta > \Delta_{\rm c}$, and
 enforce the modified conditions for the DCP in terms of
$\phi(\Delta)$:
(i) $\phi(\Delta)$ follows the same scaling law for $\Delta < \Delta_{\rm c}$ and
$\Delta > \Delta_{\rm c}$, (ii) $\phi(\Delta)$ is continuous at
$\Delta = \Delta_{\rm c}$, and (iii) 
$\frac{\partial}{\partial \Delta}\phi(\Delta)$ is discontinuous
at $\Delta = \Delta_{\rm c}$ (see
Supplemental Material for details).
With these conditions together with the assumption that the gap closes
on approaching the DCP as $|\Delta - \Delta_{\rm c}|^{\nu z}$,
one immediately
 finds: when $0 = A_- < A_+$, $z_- = d - z_+$, $\theta_{\pm} = \nu (z-z_{\pm})$, and  $\nu = 1 / (d + z - z_+)
 = 1 / (z - z_-)$, while if $A_{\pm} > 0$, $z_- = z_+$, $\theta_{\pm} =
 \nu (z - z_{\pm})$, and $\nu = 1/(d + z - z_{+})$.
Remarkably, this reduces to $\nu=1/2$
for the DCP in  XXZ model where $A_- = 0$, $A_+ > 0$ and
$z_{+} = 1$, and $\nu=1/(d+z)$
 as predicted earlier when $A_\pm > 0$ and $z_+=0$.

The  modified scaling relations 
$\hat{\xi}_{\rm t}$, $\hat{\rho}$ and 
$\varepsilon_{\rm res}$, can then be readily arrived at.
We find
\begin{eqnarray}
 \hat{\xi}_{\rm t} &\sim& t_Q^{1/(d + 2z - z_+)} , \\
 \hat{\rho} &\sim& t_Q^{-d/(d + 2z - z_+)} ,
\end{eqnarray}
and 
\begin{equation}
 \varepsilon_{\rm res}\sim t_Q^{-(d+z)/(d + 2z -z_+)} ,
\end{equation}
respectively. For the DCP at $\Delta_c$,  $\theta_+ = 1/2$, $z = 2$ and $z_+ = d = 1$,
one gets $\varepsilon_{\rm res}\sim t_Q^{-3/4}$.
We recover Eqs.~(\ref{eq:xi_t}) and (\ref{eq:epsilon1}) when $z_+=0$ which also
implies $\theta_{\pm}=\nu z $.

\begin{figure}[t]
 \begin{center}
  \includegraphics[width = 8cm]{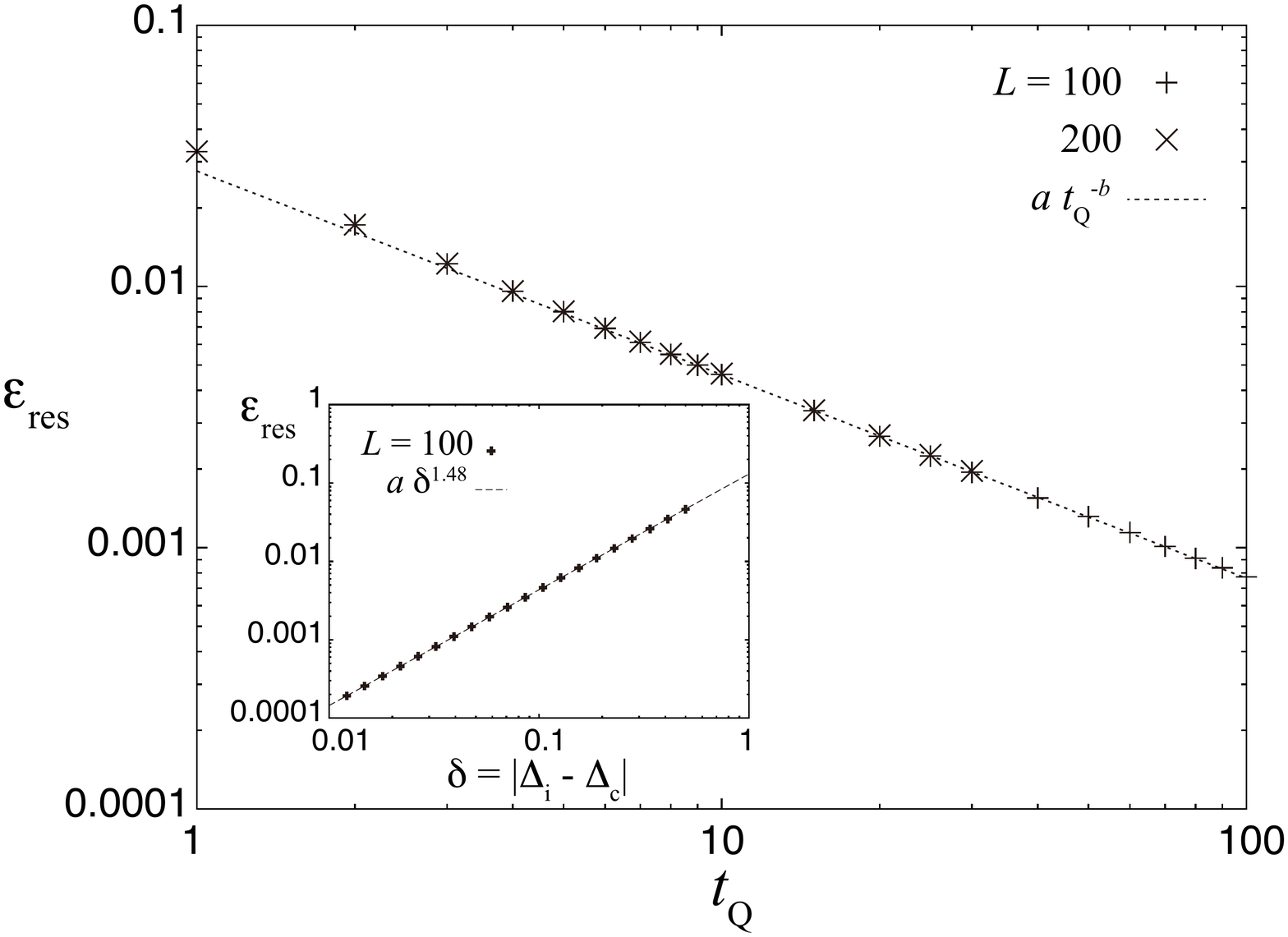}
 \end{center}
\caption{Residual energy per spin of the XXZ chain after
a temporal quench $\Delta(t) = -t/t_Q$ with time $t$
from 0 to 1 and rate $1/t_Q$ in the absence of a magnetic
field. Results of DMRG for sizes $L=100$ and $200$
and a fitting line are shown. Fitting 
is done by assuming a function $a t_Q^{-b}$ with the data of 
$L=100$ and $t_Q$'s from 10 to 100 and yields
$a = 0.0276$ and $b = 0.779$. Inset shows the scaling of $\varepsilon_{\rm res}$ for a sudden quench
of small magnitude $\delta$ starting from the gapless
 phase and ending at the DCP.}
\label{fig:SDFig3}
\end{figure}

This scaling relation is verified by numerically studying
 the time
evolution using the time-dependent
density matrix renormalization group (DMRG)
for finite systems with the open boundary condition.
We kept $m = 50$ states for each left and right blocks
in DMRG and chose the Trotter slice 
$\mathit{\Delta}t = 0.025$ \cite{comment1} with 
the initial state prepared by the static DMRG.
Figure \ref{fig:SDFig3} shows the results for sizes $L = 100$
and $200$, confirming that $\varepsilon_{\rm res}$
indeed exhibits  a  power-law scaling with an exponent 
 $-0.779$; this is consistent with the
value $-3/4$ \cite{comment2}.

If the system is quenched across  the QCP to
the gapped ferromagnetic phase, the scaling of $\varepsilon_{\rm res}$ (which is identical
to that of $\hat{\rho}$ for such a quench) satisfies 
 $\varepsilon_{\rm res}\sim t_Q^{-1/4}$; this is in perfect agreement with the result of
Pellegrini \textit{et al.} \cite{bib:Pellegrini} for a similar linear quench
of $\Delta$. For a sudden quench of small magnitude ($\delta$) from the
the gapless phase to the DCP, we find $\varepsilon_{\rm res}\sim \delta^{3/2}$ (Inset Fig.~\ref{fig:SDFig3}).
We emphasize that this is in perfect agreement with the adiabatic
perturbation theoretic prediction \cite{degrandi10} $\varepsilon_{\rm res} \sim \delta^{\nu(d+z)}$ with $\nu=1/2$ and $z=2$; this
implies that 
dynamical exponent associated with the DCP dictate the scaling of $\varepsilon_{\rm res}$.


\begin{figure}[t]
 \begin{center}
  \includegraphics[width = 8cm]{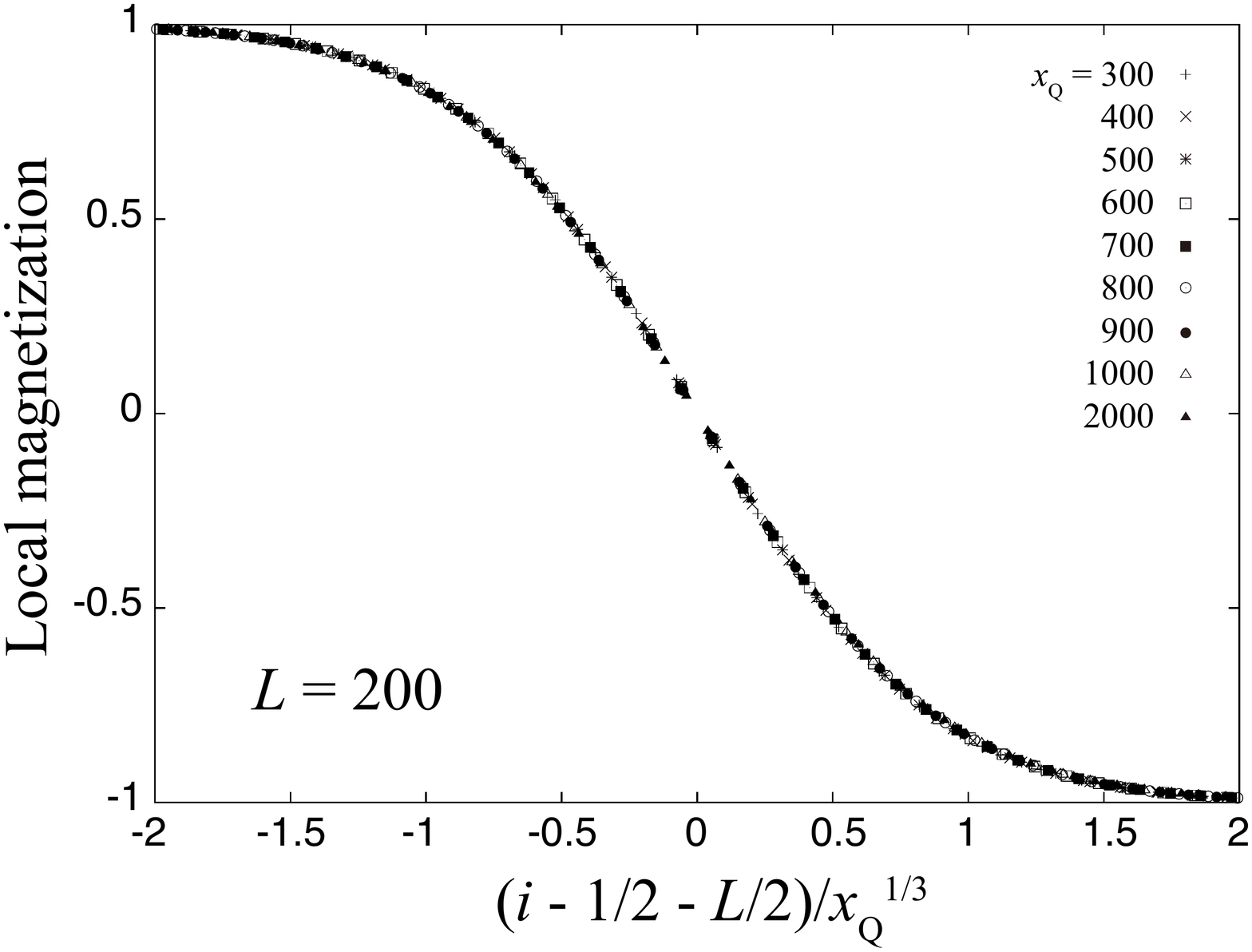}
 \end{center}
\caption{Local magnetization obtained by the static
DMRG of the XXZ chain with $\Delta = -1$
and with a sloped magnetic field $h = (i - 1/2 - L/2)/x_Q$. 
The abscissa indicates the position scaled by $x_Q^{1/3}$. 
Results for several $x_Q$ collapse into a single curve thereby confirming
the predicted scaling. }
\label{fig:SDFig2}
\end{figure}

We next move to the spatial quench with  $\Delta$ fixed to  $-1$ and
estimate $\nu_{\rm H}$.
In the homogeneous system with $h_i = h, \forall i$,
the ground state with $h = 0$ as well as $h \neq 0$
is the fully polarized state and the energy density of the 
ground state scales as $\varepsilon(h) - \varepsilon(0) = -|h|$.
The energy gap, on the other hand, scales with $h$ as
$\mathit{\Delta}E \sim |h| \sim |h|^{\nu_{\rm H} z}$
\cite{comment3}; with $z=2$, this
immediately implies $\nu_{\rm H} =1/2$. 
We emphasize here that 
this value of $\nu_{\rm H}$ differs from $1/(d + z)$ predicted before.
This is due to the  fact that the ground state is completely spin polarized without fluctuations resulting 
in a $\xi_{\rm H}$ that is
independent of $h$ and always $\sim 1$. Consequently, the the ground-state
energy density does no longer scale as 
$\sim |h|^{(d + z)\nu_{\rm H}}$, rather scales as $|h|^{z\nu_{\rm H}}$; 
using this scaling one arrives at $\nu_{\rm H} = 1/z$
and $\hat \xi_{\rm s} \sim x_Q^{1/(z+1)}=x_Q^{1/3}$ with $z=2$.

Let us now
apply a sloped field $h_i = (i-1/2-L/2)/x_Q$
choosing an even  $L$ and probe the local
magnetization,
 $m_i = \langle \sigma_i^z\rangle$ of the ground state. The scaling theory
presented above predicts that the characteristic length scale
near $i = (L+1)/2$ is  given by $\hat{\xi}_s\sim
x_Q^{1/3}$.  
In~fact, one then expects the scaling $m_i \sim {\cal F}\left(\frac{i-\frac{1}{2}-\frac{L}{2}}{x_Q^{1/3}}\right)$, where ${\cal F}$ is
the scaling function;
remarkably this prediction  perfectly matches with the numerical results obtained using
 static DMRG as presented in Fig.~\ref{fig:SDFig2}.


In conclusion, we provided  universal scaling relations
for quenches in the vicinity
of a quantum  DCP  for both spatial and temporal quenches. Furthermore,
we have established that indeed there exists a DCP in the phase
diagram of a spin-1/2 XXZ chain. The critical exponents and hence the scaling relations
following quenches get non-trivially modified since this DCP occurs at the
terminating point of a gapless critical line. Most importantly, we  posit  generic
scaling relations which perfectly reduce to both the situations (Ising and XXZ)  considered here.
We emphasize that the scaling predictions  for the XXZ chain are  verified
numerically.
An experimental demonstration of our results is
expected in a cold atomic system in an optical lattice
or a trapped-ion system.
%
%
%

SS thanks MEXT, Japan for support through
grant No. 26400402.  We acknowledge Jun-ichi Inoue and Shraddha Sharma for discussions.


\vspace{-\baselineskip}


\pagebreak

%
%
\widetext
\begin{center}
\textbf{\large Supplemental Material on ``Universal scaling in quenches across a discontinuity critical point''}\\
\vspace{0.5cm}
{Sei Suzuki$^1$ and Amit Dutta$^2$}\\
$^1$ Department of Liberal Arts, Saitama Medical
University, Moroyama, Saitama 350-0495, Japan\\
$^2$Department of Physics, Indian Institute of
Technology, Kanpur 208 016, India
\end{center}

\newcommand\bea{\begin{eqnarray}}
\newcommand\eea{\end{eqnarray}}
\newcommand\beq{\begin{equation}}
\newcommand\eeq{\end{equation}}
\newcommand\bi{\bibitem}
\newcommand{\ct}{\cite}

\def\noi{\noindent}
\def\non{\nonumber}
\def\dag{\dagger}
\def\al{\alpha}
\def\om{\omega}
\def\de{\delta}
\def\De{\Delta}
\def\ep{\epsilon}
\def\ga{\Delta}
\def\ka{\kappa}
\def\la{\lambda}
\def\si{\sigma}
\setcounter{equation}{0}
\setcounter{figure}{0}
\setcounter{table}{0}
\setcounter{page}{1}
\makeatletter
\renewcommand{\theequation}{S\arabic{equation}}
\renewcommand{\thefigure}{S\arabic{figure}}
\renewcommand{\bibnumfmt}[1]{[S#1]}
\renewcommand{\@cite}[1]{[S#1]}
\makeatother

\begin{figure}[h]
 \begin{center}
  \includegraphics[width = 8cm]{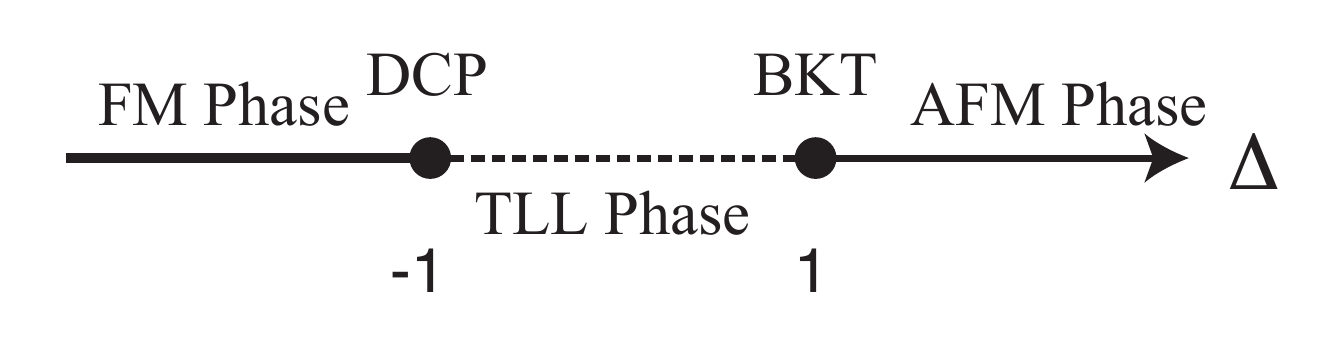}
 \end{center}
\caption{The phase diagram of Hamiltonian \eqref{eq:suppl:H_XXZ}; the gapless
 region lying between $-1 <  \Delta \leq 1$ is the
Tomonaga-Luttinger liquid (TLL) phase. The transition between the gapless phase and the antiferromagnetically (AFM) ordered
phase at $\De=1$ is Berezinski-Kosterlitz-Thouless (BKT) transition point while the critical point separating the gapless phase
from the ferromagnetic (FM) phase at $\Delta=-1$ is a
discontinuity critical point (DCP).}
\label{fig:SupplXXZ}
\end{figure}

Here, we shall establish that the quantum critical point appearing at the boundary of the gapless critical line and
the gapped ferromagnetic phase of the spin-1/2 XXZ chain (Hamiltonian (9) of the main text) with $h_i=0, \forall i$ ,
\begin{equation}
 H = \sum_{i=1}^{L-1}
  \left(\sigma_i^x \sigma_{i+1}^x + \sigma_i^y \sigma_{i+1}^y
   + \Delta\sigma_i^x\sigma_{i+1}^z \right)
\label{eq:suppl:H_XXZ}
\end{equation}
 at $\Delta=\Delta_c=-1$ is
indeed a discontinuity critical point (DCP) \ct{bib:suppl:fisher82,bib:suppl:nishimori11}
and show how the associated critical exponents get altered in
this continuous symmetry model where the DCP is located at the boundary of a gapless critical line; in the process, we
shall also touch upon the scaling relations and critical exponents associated with a generic DCP. The complete phase diagram of the chain
is shown in Fig.
\ref{fig:SupplXXZ} \ct{bib:suppl:Giamarchi}.

 To this end, we recall that
the ground-state energy density $\varepsilon(\Delta)$
in the  $\sigma_{\rm tot}^z = 0$ manifold 
is given exactly by \cite{bib:suppl:yang66}
\begin{equation}
 \varepsilon(\Delta) 
 = \left\{
 \begin{array}{cl}
  \Delta & (\mbox{for $\Delta < -1$})\\
\\
  \displaystyle \cos \mu - \sin^2\mu\int_{-\infty}^{\infty}
   \frac{dx}{\cosh\pi x (\cosh 2\mu x - \cos \mu)}
   & (\mbox{for $-1 < \Delta < 1$}) 
 \end{array}
\right. ,
\end{equation}
where $\mu$ is defined through the relation $\cos \mu = \Delta$.
When $0 < \Delta + 1 \ll 1$, making an expansion
$\cos\mu \approx -1 + \frac{1}{2}\delta^2$ 
(which implies $\frac{1}{2}\delta^2 \approx \Delta + 1$)
and  retaining lowest order terms in Taylor expansion in $\delta$, we arrive at
\begin{eqnarray*}
 \varepsilon(\Delta) &\approx&
  - 1 - \frac{1}{3\pi}\delta^3 + \mathcal{O}(\delta^4) \\
 &=& -1 + \frac{2\sqrt{2}}{3\pi}(\Delta + 1)^{3/2}
  + \mathcal{O}\left((\Delta+1)^2\right) .
\end{eqnarray*}
Consequently, one gets
\begin{equation}
 \varepsilon(\Delta) - \varepsilon(\Delta_{\rm c})
  \approx
  \left\{
   \begin{array}{cl}
    \Delta - \Delta_{\rm c} & (\mbox{for $\Delta < -1$})\\
    \\
    \displaystyle \frac{2\sqrt{2}}{3\pi}(\Delta - \Delta_{\rm c})^{3/2}
     & (\mbox{for $-1 < \Delta < 1$}) 
   \end{array}
       \right. ,
\end{equation}
where we have used $\Delta_{\rm c} = -1.$
This immediately reveals that
the derivative of $\varepsilon(\Delta) - \varepsilon(\Delta_c)$
 is discontinuous at $\Delta = \Delta_c$, and
 establishes that this point is a DCP.
 
 Now, in order to settle the value of $\nu$, we compare
 $\varepsilon(\Delta) - \varepsilon(\Delta_c) \sim (\Delta - \Delta_c)^{3/2}$
 for $\Delta > \Delta_c$ with the conventional scaling of the ground state energy density\ct{bib:suppl:Sachdev,bib:suppl:suzuki13}, $(\Delta - \Delta_c)^{(d + z)\nu}$ . This comparison leads us to the relation $(d+z)\nu = 3/2$.
Note here $z = 2$, since the spin wave accounts for the excitation
at $\Delta = \Delta_{\rm c}$ even in the $\sigma_{\rm tot}^z = 0$
manifold
and the spectrum is quadratic in the momentum.
Therefore, taking $d = 1$ into account, one finds $\nu = 1/2$.
Surprisingly, this is inconsistent with  the prediction $1/(d + z) = 1/3$.
As we shall elaborate below, this anomaly emerges  from the existence of a critical line for
$\Delta > \Delta_{\rm c}$.

To modify the scaling theory accordingly, we assume that
the spectrum follows the form 
$|\Delta - \Delta_c|^{\theta_{\pm}}k^{z_{\pm}} + k^z$, where 
$(\theta_-, z_-)$ and $(\theta_+, z_+)$ represent exponents on the lower-$\Delta$ 
and the higher-$\Delta$ sides of the DCP, respectively.
We here demand $0 \leq z_{\pm}< z$.  Clearly, one
can readily  retrieve the quadratic scaling of the gap right  at the DCP by setting $\Delta=\Delta_c$.
Comparison of the spectrum with the conventional scaling form $|\Delta - \Delta_{\rm c}|^{\nu z}$ 
yields
\begin{equation}
 \theta_{\pm} = \nu (z - z_{\pm}) . 
\label{eq:Suppl:theta}
\end{equation}
Next, we assume that the correlation length
has the form $\xi(\Delta) \sim 1 + A_{\pm}|\Delta - \Delta_{\rm c}|^{-\nu}$,
where $\pm$ corresponds to the sign of $\Delta - \Delta_{\rm c}$.
It is to be  noted  that $\xi(\Delta) \sim 1$ when $A_{-} = 0$ (or $A_+ = 0$);
this situation arises when the fluctuation of the order parameter
is entirely absent which indeed happens, as we shall show below, in the   ground state of the XXZ chain for $\Delta < \Delta_{\rm c}$.
Having assumed the the scaling form of the spectrum and the correlation length,
the ground-state energy density can be written as
\begin{equation}
 \varepsilon(\Delta) - \varepsilon(\Delta_{\rm c})
  \sim |\Delta - \Delta_{\rm c}|^{\theta_{\pm} + \nu
  z_{\pm}}\xi(\Delta)^{-d}
  \sim |\Delta - \Delta_{\rm c}|^{\theta_{\pm} + \nu z_{\pm}}
  (1 + A_{\pm}|\Delta - \Delta|^{-\nu})^{-d} .
\label{eq:Suppl:gse}
\end{equation}
We emphasize that, in general, this quantity does not scale with an
identical exponent on the lower- and higher-$\Delta$ sides. To resolve this
asymmetry,
let us  introduce the new quantity:
\begin{equation}
 \varphi (\Delta) = \begin{cases}
   (\varepsilon(\Delta) - \varepsilon(\Delta_{\rm c}) |\Delta -
    \Delta_{\rm c}|^{\nu z_-}) & \mbox{for $\Delta < \Delta_{\rm c}$} \\
   (\varepsilon(\Delta) - \varepsilon(\Delta_{\rm c}) |\Delta -
    \Delta_{\rm c}|^{\nu z_+}) & \mbox{for $\Delta > \Delta_{\rm c}$}
		     \end{cases},
\label{eq:Suppl:phi_def}
\end{equation}
and demand modified conditions of the DCP as follow:
(i) this quantity follows the same scaling relation on both sides
of the critical point 
and that (ii) it is continuous while (iii) its derivative is 
discontinuous at the critical point:
\begin{eqnarray}
&({\rm i})&~~~ \varphi(\Delta_{\rm c} - \epsilon) \sim
 \varphi(\Delta_{\rm c} + \epsilon) \sim |\epsilon|^a
 ~~~ \mbox{with an $a > 0$}, \\
\nonumber\\
&({\rm ii})&~~~ \varphi_-(\Delta_{\rm c} - 0) = \varphi_+(\Delta_{\rm c} + 0)
 ,\\
\nonumber\\
&({\rm iii})&~~~ \frac{\partial \varphi_-(\Delta_{\rm c} - 0)}{\partial \Delta}
  \neq \frac{\partial \varphi_+(\Delta_{\rm c} + 0)}{\partial \Delta} .
\end{eqnarray}
We hereafter discuss the cases with $A_{-} = 0$ and $A_- \neq 0$
separately, with $A_+\neq 0$ in both the situations.

We first consider the situation where $A_- = 0$. Equations~(\ref{eq:Suppl:gse}) and (\ref{eq:Suppl:phi_def}) then yield
\begin{equation}
 \varphi (\Delta) \sim
  \begin{cases}
   (\Delta_{\rm c} - \Delta)^{\theta_-} & \mbox{for $\Delta < \Delta_{\rm
   c}$} \\
   (\Delta - \Delta_{\rm c})^{\theta_+ + d\nu} & \mbox{for $\Delta >
   \Delta_{\rm c}$}
   \end{cases},
\end{equation}
Using Eq.~(\ref{eq:Suppl:theta}) and the {condition} (i) on
$\varphi(\Delta)$,
one obtains $(z - z_-)\nu = (d + z - z_+)\nu$, which immediately
reduces to $z_- = z_+ - d$. The condition (ii) along with (i) result in  constraint relations,
$z_- \leq z$ and $z_+ \leq d + z$  (though these are always satisfied under
the restriction we have imposed on $z_{\pm}$).
  Finally, by the condition (iii), one
obtains
$\nu = 1/(d + z - z_+) = 1/(z - z_-)$.
It is noteworthy that we are now left with only two independent exponents out of the six exponents
$z$, $\nu$, $z_{\pm}$, and $\theta_{\pm}$ involved in the scaling. For instance,
if $z$ and $z_+$ are known for a given $d$, the other exponents are automatically
fixed.
Now let us return to the XXZ chain. 
The ground-state energy density and the energy gap 
of the chain scale identically as $|\Delta - \Delta_{\rm c}|$
for $\Delta < \Delta_{\rm c}$. This scaling together with the 
the fact that the spectrum is  gapped in this region imply $A_- = 0$ and establish that
the  XXZ chain indeed corresponds to the present situation.
Applying the value $d = z_+ = 1$ and $z = 2$ of the XXZ chain to
the generic relations, one obtains $\nu = 1/2$, $z_- = 0$, $\theta_- = 1$,
and $\theta_+ = 1/2$, which are completely consistent with the
property of the XXZ chain.

Next, we consider the case with $A_- \neq 0$. Equations (\ref{eq:Suppl:gse}) and
(\ref{eq:Suppl:phi_def})  now reduce to
\begin{equation}
 \varphi (\Delta) \sim \begin{cases}
    (\Delta_{\rm c} - \Delta)^{\theta_- + d\nu} & 
			\mbox{for $\Delta < \Delta_{\rm c}$} \\
    (\Delta - \Delta_{\rm c})^{\theta_+ + d\nu} &
			\mbox{for $\Delta > \Delta_{\rm c}$} 
			\end{cases} .
\end{equation}
Following the same set of arguments as in the previous situation results
in $z_- = z_+$ and $\nu = 1/(d + z - z_+) = 1/(d + z - z_-)$.
If one sets $z_+ = 0$, one obtains
$\nu = 1/(d + z)$. Hence the
correlation length exponent of a generic isolated DCP, mentioned
in the main text, is retrieved.

To summarize, we have provided the scaling theory for a generic DCP
based on the conditions mentioned above.
When a DCP characterized by the dynamical exponent $z$ ($> 0$)
is located between two critical phases with different dynamical
exponents $z_-$ and $z_+$ (assuming $0 \leq z_- < z_+ < z$ without loss of generality),
the exponents $z_-$ and $z_+$ are constrained by $z_+ - z_- = d$ and
the correlation length exponent is given by 
$\nu = 1/(d + z - z_+) = 1/(z - z_-)$. Moreover,
the critical amplitude of the correlation length satisfies the condition
$A_-/A_+ = 0$. 
Note that these rules are valid even when one of the two phases
is gapped ($z_- = 0$).
On the other hand, when a DCP with $z>0$ is located between two critical phases
with an identical exponent $z_- = z_+ = z' < z$ or two gapped phases
($z_- = z_+ = z' = 0$), 
one finds
$\nu = 1/(d + z - z')$ and the critical amplitudes of the correlation
length are finite on the both sides of the DCP.


\begin{thebibliography}{99}
\bibitem{comment1} We have confirmed this choice reduces
the error within $\sim 10^{-6}$.
\bibitem{comment2} We attribute the small difference to the finite-size effect and
the small values  of $t_Q$.
\bibitem{comment3} It is to be noted that the the ground state and the energy gap here are 
defined in the whole manifold of the spin states.
\end{thebibliography}
\end{document}